\documentclass[12pt,amsfonts,a4paper]{article}    %
\usepackage{amsfonts}

\def\one{1\hskip-.37em 1}

\def\half{\textstyle{\frac{1}{2}}}

\def\l{\lambda}

\def\tint{{\textstyle\int}}

\def\s{\hskip.08em}

\def\b{\begin{eqnarray*}}  %takes no eqn numbers
\def\e{\end{eqnarray*}}    %takes no eqn numbers
\def\bn{\begin{eqnarray}}  %takes eqn numbers
\def\en{\end{eqnarray}}   %takes eqn numbers

\usepackage{color}

\def\<{\langle}
\def\>{\rangle}

\def\no{\nonumber}

\def\{{\lbrace}
\def\}{\rbrace}
\title{A New Rule for Quantization \\that Resolves All Problems}     %%%  FOR DICE2018
\author{John R. Klauder\\  %footnote{john.klauder@gmail.com}\\
Department of Physics and Department of Mathematics\\
University of Florida,
Gainesville, FL 32611-8440, U.S.A.\\
john.klauder@gmail.com }
\date{ }
\begin{document}
\maketitle

\begin{abstract} While canonical quantization solves many problems there are some problems where it fails. A close examination of the
classical/quantum connection leads to a new connection that permits quantum and classical realms to coexist, as is the case in the real world.
For those problems for which conventional quantization works well, the new procedures yield identical results. However, we offer an
example that fails when quantized conventionally but succeeds when quantized with the new procedures.\end{abstract}

\section{Canonical Quantization vs. \\Enhanced Quantization}
 For a single degree of freedom, canonical quantization (CQ) takes classical phase space variables, $(p,q)$, and promotes them into Hilbert space operators, $(P,Q)$.
 %These basic operators obey the commutator $[Q,P]=i\hbar \one$, an analog of the classical Poisson bracket $\{q,p\}=1$.
  The classical Hamiltonian  $H(p,q)$ is promoted to the quantum Hamiltonian $ {\cal{H}}(P,Q)$, specifically
 ${{\cal{H}}}(P,Q)=H(P,Q)+{\cal{O}}(\hbar; P, Q)$  provided the
 phase space variables $p$ and $q$ are ``Cartesian coordinates'' \cite{D} (footnote, page 114), even though phase space has no metric to determine
 Cartesian coordinates.

Enhanced quantization (EQ) features a different classical/quantum connection story. It starts by introducing self-adjoint operators $P$ and $Q$.
%that satisfy the commutator $[Q,P]=i\hbar \one$.
The quantum Hamiltonian ${\cal{H}}(P,Q)$ is chosen as a self-adjoint operator built from the basic variables.
Schr\"odinger's equation follows from the action functional    %
\bn A_Q=\tint_0^T \<\psi (t)|[ i\hbar(\partial/\partial t) - {{\cal{H}}}(P,Q)] |\psi (t)\> \,  dt \en   %\end{document}
 by stationary variations of the normalized Hilbert space vectors $|\psi (t)\>$. However, a {\it macroscopic} observer can only vary a subset of vectors of a {\it microscopic} system such as
 \bn  |p,q\> \equiv e^{-iqP/\hbar}\s e^{ipQ/\hbar} |0\>\;, \en    where
   $(p,q)\in \mathbb{R}^2$, and the normalized fiducial vector $|0\>$ is chosen as a solution of the equation $(\omega Q +i P ) |0\>=0$.
   %Recall that Galilean invariance permits a change of the system by position or velocity (note: $v=p/m$ for fixed $m$) by a change of the observer {\it %without disturbing the system}. All vectors $|p ,q\>$ are normalized as well because $P$ and $Q$ are self adjoint.
   The reduced (R) quantum action functional is given by
 \bn A_{Q(R)}=\tint_0^T \<p(t),q(t)|[i\hbar(\partial/\partial t)-{\cal{H}}(P,Q)]|p(t),q(t)\>  \;  dt \;,   \en
   which leads to
  \bn A_{Q(R)}=\tint_0^T [p(t) \dot{q}(t) - H(p(t),q(t))]\;  dt\;, \en
 a natural candidate to be the classical Hamiltonian, with
 $\hbar>0$, as it is in the real world! In this expression
 \bn  && \hskip-3.4em H(p,q)\equiv  \< p,q| {\cal{H}}(P, Q) |p,q\> \no\\
       &&= \<0| {\cal{H}}(P+p\one, Q+q\one) |0\>  \no \\
      &&= {\cal{H}}(p,q)+{\cal{O}}(\hbar;p,q) \;. \en
   Apart from possible $\hbar$ corrections, ${\cal{H}}(P,Q)=H(P,Q)$, which {\it exactly}  represents the goal of seeking Cartesian coordinates. Although phase space has no metric to identify Cartesian coordinates, Hilbert space has one. It follows that a scaled form of the Fubini-Study metric \cite{FS} for ray vectors, specifically $d\sigma^2\equiv 2\hbar\s [\s \|\s d|p,q\>\|^2-|\<p,q|\,d|p.q\>|^2]$, yields the result $ d\sigma^2=\omega^{-1} dp^2+\omega\s dq^2$, which   indeed implies Cartesian coordinates for a flat phase space.  Consequently, EQ can lead to the same good results given by CQ, but EQ can do even more. % and so we claim that $CQ\subset EQ$.

        In EQ the classical and quantum Hamiltonian are related by the weak correspondence principle \cite{DD},
    \bn H(p, q)= \<p, q | {\cal{ H}}(P, Q) | p, q \>\;, \label{e3} \en
 which holds for self-adjoint canonical variables.
 In CQ the canonical operators $P$ and $Q$ are required to be irreducible, and in EQ they may also be irreducible. However, from (\ref{e3})
         it follows that the
         quantum variables may also be reducible.  The following problem can be properly solved with reducible operators, but cannot be properly solved with irreducible  operators.
         %This feature may seem foreign to the reader, but perhaps an example of the power of EQ may help the reader to overcome any doubts. START

\section{Rotationally Symmetric Models}
This example involves many variables, and it has been examined before \cite{ka, BCQ}; here we only offer an overview of the solution. The classical Hamiltonian is given by
\bn H(p, q)=\half \Sigma_{n=1}^N [p_n^2+m_0^2 q_n^2]
+\l_0 \{\Sigma_{n= 1}^N q_n^2 \}^2 \;, \en
where $N\le \infty$, and the Poisson bracket is $\{q_k, p_l\}=\delta_{kl}$.
     For $\l_0>0$ and  $N=\infty$ the quantum model is trivial (i.e., Gaussian) when CQ and irreducible operators are used
     %(if you have doubts, try to solve it; or see \cite{ka,BCQ} for a proof).
     To obtain reducible operators we introduce a second  set of canonical operators $\{R_n,S_n\}$
      which commute with all $\{P_n,  Q_n\}$. Triviality of such models forces us to restrict the fiducial vector to be a Gaussian in a Schr\"odinger representation,
     i.e., where $Q_n\s|x\>=x_n\s |x\>$. Thus $(m_0 Q_n+i P_n) |0\>=0$ for the free model, with the property that
     \bn (m_0 Q_n-i P_n) \cdot(m_0 Q_n+i P_n)&&\hskip-1.3em  =P_n^2+m_0^2\s Q_n^2+i\s m_0\s [Q_n,P_n] \no \\
               &&\hskip-1.25em  =\, :\s P_n^2+m_0^2\s Q_n^2\s :\;, \en
     for all $n$, $1\le n\le N$, with normal ordering denoted by :$\s(\cdot)\s$:. It follows, for example, that
        \bn \<p,q|\, \half \,\Sigma_{n=1}^N \s( P_n^2+m_0^2\s Q_n^2)\s: + w\,:\s \{\Sigma_{n=1}^N \,( P_n^2+m_0^2\s Q_n^2)\s \}^2:\,|p,q\> \no  \\
             = \half\,\Sigma_{n=1}^N \s (p^2_n+m_0^2\s q^2_n)+w \{\Sigma_{n=1}^N (p^2_n+m^2_0 \s g^2_n)\s\}^2\;, \en
             which offers a clue of how to proceed.

     For the interacting model with reducible operators this example
     is changed to $[m(Q_n+\zeta S_n)+i P_n] |\zeta\>=0$ and
$[m(S_n+\zeta Q_n)+ i R_n ] |\zeta\>=0$, for all $n, \, 1\le n\le N$, where $ 0<\zeta< 1$. We next introduce new reduced states as
   \bn |p,q;\zeta\>\equiv e^{-i\Sigma_{n=1}^N q_n P_n/\hbar}\s e^{i\Sigma_{n=1}^N p_n Q_n/\hbar} |\zeta\> \;, \en
which still span the relevant Hilbert space. Then it follows  that
\bn   &&\hskip-3.4em H( p, q)=\<p, q;\zeta| \,\half \, \{: \Sigma_{n=1}^N [P_n^2+m^2(Q_n+\zeta S_n)^2] :\no \\
    &&\hskip6em + : \Sigma_{n=1}^N [R_n^2+m^2(S_n+\zeta Q_ n)^2] : \} \no \\
   &&\hskip6em  + v : \{ \Sigma_ {n=1}^N [R_n^2+m^2(S_n+\zeta Q_n)^2] \}^2 : |p,q;\zeta\> \no \\
    &&=\half\, \Sigma_{n=1}^N [p_n^2+m^2 (1+\zeta^2) q_n^2] +
      v\s \zeta^4 m^4 \{ \Sigma_{n=1}^N q_n^ 2 \}^2 \no \\
  && \equiv \half\, \Sigma_{n=1}^N [p_n^2+ m_0^2 q_n^2]
  + \l_0 \{ \Sigma_{n=1}^N q_n^2 \}^2 \en
  as desired!

  Further discussion of these models is available in \cite{ka,BCQ}.

\section{Additional Enhanced\\ Quantization Examples }
Enhanced quantization has also been applied to other problems, among which ultralocal scalar models \cite{ul, BCQ} admit complete solutions and
additionally suggest proposals for more complicated examples. These
include covariant scalar fields \cite{E3,E1,E2,PP,JK1},
 and quantum gravity \cite{E3,G1,G2,G3}.    %which exploits irreducible affine fields.
 The use of special variables to discuss idealized
 cosmological models \cite{Y1,WW} has led to gravitational bounces rather than an initial singularity of the universe. Further use of
 special variables as applied to idealized gravitational models has been given in various papers, e.g., \cite{T1,T2,T3}, and
 references therein.

\end{document}